\documentclass[12pt,a4paper]{article}
\usepackage{amsfonts}
\usepackage{a4wide}
\usepackage{amssymb}
\usepackage{latexsym}
\usepackage{amssymb}
\usepackage{amsmath}
\usepackage{bbm}
\usepackage{ifpdf}
\ifpdf
\usepackage[pdftex]{graphicx}
\usepackage[pdftex,unicode,implicit]{hyperref}
\hypersetup{%
  pdftitle    = {Supersymmetry  and cosmic censorship.},
  pdfkeywords = {N=2 Supersymmetry, attractors, BPS black holes},
  pdfauthor   = {T. Ortín},
  pdfcreator  = {pdf\LaTeXe\ with package \flqq hyperref\frqq},
  pdfproducer = {pdf\LaTeXe\ with package \flqq hyperref\frqq},
  pdfpagemode = None,  
  pdffitwindow= true,  
  unicode     = true,
  plainpages  = false,
  colorlinks  = true,  
  citecolor   = blue,
  urlcolor    = red,
  linkcolor   = green
}
\newcommand{\hepth}[1]{arXiv:{\tt \href{http://www.arXiv.org/abs/hep-th/#1}{hep-th/#1}}}

\else
  \usepackage[dvips]{graphicx}
  \usepackage[unicode,implicit]{hyperref}
  \newcommand{\hepth}[1]{arXiv:{\tt hep-th/#1}}

\fi
\begin{document}
\begin{flushright}
\small
IFT-UAM/CSIC-06-53\\
{\bf hep-th/0611117}\\
November $10^{\rm th}$, $2006$
\normalsize
\end{flushright}
\begin{center}
\vspace{2cm}
{\LARGE {\bf Supersymmetry and cosmic censorship}}\footnote{To be published in
  the Proceedings of the $2^{\rm nd}$ Workshop and Midterm Meeting of the
  Marie Curie Research Training Network \textsl{Constituents, Fundamental
    Forces and Symmetries of the Universe}, Naples, Italy, October 9-13 2006.
} \vspace{2cm}

{\sl\large Tom{\'a}s Ort\'{\i}n}
\footnote{E-mail: {\tt Tomas.Ortin@cern.ch}}

\vspace{1cm}

{\it Instituto de F\'{\i}sica Te\'orica UAM/CSIC\\
Facultad de Ciencias C-XVI,  C.U.~Cantoblanco,  E-28049-Madrid, Spain}\\

\vspace{2cm}


{\bf Abstract}

\end{center}

\begin{quotation}\small
  We show that requiring unbroken supersymmetry \textit{everywhere} in
  black-hole-type solutions of $N=2,d=4$ supergravity coupled to vector
  supermultiplets ensures in most cases absence of naked singularities.  We
  show that the requirement of global supersymmetry implies the absence of
  sources for NUT charge, angular momentum, scalar hair and negative energy,
  for which there is no microscopic interpretation in String Theory.  These
  conditions exclude, for instance, singular solutions such as the Kerr-Newman
  with $M=|q|$, which fails to be everywhere supersymmetric. There are,
  nevertheless, everywhere supersymmetric solutions with \textit{global}
  angular momentum and non-trivial scalar fields. 
  
  We also present similar preliminary results in $N=1,d=5$ supergravity
  coupled to vector multiplets.
\end{quotation}

\newpage

\pagestyle{plain}

\section{Introduction: the 1992 SUSY \textit{versus} cosmic censorship 
conjecture}

A possible relation between cosmic censorship and supersymmetry follows from
the observation that in the simplest supergravity theories the BPS bound
coincides with the condition of existence of event horizons in static
black-hole-type solutions.  Thus, for the Schwarzschild solution, which is a
solution of pure $N=1,d=4$ supergravity, the BPS bound $M\geq 0$ ensures the
existence of a regular horizon. When the bound is saturated $M=0$, there is
unbroken supersymmetry and a regular solution: Minkowski spacetime.

The Reissner-Nordstr\"om solution
\begin{equation}
ds^{2}  = 
{\displaystyle\frac{(r-{r_{+}})(r-{r_{-}})}{r^{2}}}dt^{2} 
-{\displaystyle\frac{r^{2}}{(r-{r_{+}})(r-{r_{-}})}}dr^{2}
-r^{2}d\Omega^{2}_{(2)} \, ,\hspace{.6cm} 
r_{\pm} =  M\pm \sqrt{M^{2}-q^{2}}\, ,
\end{equation}
is a solution of pure $N=2,d=4$ supergravity and has a regular event horizon
for $M^{2}\geq q^{2}$, which is the BPS bound. When the bound is saturated,
there is unbroken supersymmetry and a regular solution: the extreme
Reissner-Nordstr\"om black hole. 

It was then proposed that supersymmetry works as a cosmic censor
\cite{Kallosh:1992ii}. The conjecture is supported by the relation between
these two concepts and the positivity of the energy.

It was, however, quickly realized that the conjecture fails for the simplest
black-hole-type stationary supersymmetric solutions of pure $N=2,d=4$
supergravity, which have the general form
\cite{Perjes:1971gv,kn:IW,Tod:1983pm}
\begin{equation}
  \begin{array}{rcl}
ds^{2} & = & |{V}|^{2}(dt +{\omega})^{2} 
-|{V}|^{-2}d\vec{x}^{2}\, , \\
& & \\
\text{where} & &  
d{\omega} = i\star_{(3)}
|{V}|^{-2}[\bar{V}^{-1}
dV^{-1}-V^{-1}d\bar{V}^{-1}]\, ,\,\,\,\,
\text{and}\,\,\,\,
\nabla^{2}_{(3)}V^{-1} = 0\, .\\
\end{array}
\end{equation}
For instance, the extreme Reissner-Nordstr\"om-Taub-NUT solution
\cite{kn:Bri}, which is characterized by the complex harmonic function
\begin{equation}
V^{-1}= 1+ \frac{M+iN}{r}\,\,\,\, \Rightarrow\,\,\, 
\omega = 2N \cos{\theta}d\phi\, ,
\end{equation}
has wire singularities at $\theta=0,\pi$.  They are always associated to the
integral 
\begin{equation}
\int_{\partial\Sigma^{3}}d{\omega} = \int_{\Sigma^{3}}d^{2}{\omega}=
-8\pi {N}\, ,  
\end{equation}
and can be removed, but only at the expense of asymptotic flatness
\cite{kn:M}. 

Another example is provided by the supersymmetric ($M=|q|$) Kerr-Newman
solution \cite{Kerr:1963ud,kn:NCCEPT} $J=M\alpha$, characterized by
\begin{equation}
V^{-1} = 1 +\frac{M}{\sqrt{x^{2}+y^{2}+(z-i\alpha)^{2}}}\, ,
\end{equation}
which has a naked singularity in the ring $x^{2}+y^{2}= \alpha^{2}\, ,\,\,\,
z=0$. 

Finally, the multipole solutions of this theory
\begin{equation}
\label{eq:multipole}
V^{-1}= 1+ \sum_{n}\frac{M_{n}+iN_{n}}{|\vec{x}-\vec{x}_{n}|}\, ,
\end{equation}
always seem have wire singularities associated to the points at which
$d^{2}{\omega}\neq 0$, which can be seen as sources of NUT charges $N_{n}$,
and there seems to be no choice of $M_{n},N_{n},\vec{x}_{n}$ that eliminates
them. It was then conjectured by Hartle and Hawking \cite{Hartle:1972ya} that
the only regular black-hole solutions of this class were the
Papapetrou-Majumdar \cite{kn:Pa,Majumdar:1947eu} solutions ($N_{n}=0\,
,\,\,\,\forall n$) describing static extreme Reissner-Nordstr\"om black holes
in equilibrium \cite{kn:BrilLin}.

Observe that, coincidentally, no microscopic String Theory descriptions of NUT
charge or angular momentum (preserving supersymmetry) seem to exist.
    
We will show that, actually, only the regular solutions which can be described
by String Theory are truly supersymmetric \textit{everywhere}.


\section{The 2006 SUSY \textit{versus} cosmic censorship conjecture}

The explicit knowledge of the most general supersymmetric black-hole-type
solutions of $N=2,d=4$ SUGRA theories \cite{Behrndt:1997ny,Meessen:2006tu} has
led as to reformulate the 1992 conjecture extending the requirement of
supersymmetry to the \textit{sources} that give rise to the macroscopic fields
\cite{Bellorin:2006xr}. The conjecture says now that supersymmetric,
asymptotically-flat, black-hole-type solutions satisfying the following
conditions will be regular black holes without naked singularities:
\begin{itemize}
\item[\textbf{I}] The solutions must be {\it everywhere} supersymmetric. In
  particular this implies that 
  \begin{enumerate}
  \item The integrability conditions of the Killing spinor equations (KSIs
    \cite{Kallosh:1993wx,Bellorin:2005hy}) must be satisfied
    {\textit{everywhere}}. We will see that this always requires
    $d^{2}{\omega}=0$ {\textit{everywhere}}.
  \item The masses of each of the sources of the solutions should be positive.
  \end{enumerate}
\item[\textbf{II}] In presence of scalars, the attractor equations
  \cite{Ferrara:1995ih,Strominger:1996kf,Ferrara:1996dd,Ferrara:1996um}
\begin{displaymath}
\left. \mathfrak{D}_{i}\mathcal{ Z}\right|_{Z^{i}=Z^{i}_{\rm fix}}=0\, .  
\end{displaymath}
must be satisfied at each of the sources for admissible values of the scalars
(no hair) and the value of the central charge must be finite.

\end{itemize}

These conditions should be enough to ensure the finiteness and positivity of
$-g_{rr}$ everywhere. Now let us see how these conditions select regular
solutions that can be described microscopically by String Theory.


\subsection{Pure $N=2,d=4$ SUGRA} 

The KSIs are relations between the equations of motion of the bosonic fields
$\mathcal{E}^{\mu\nu}\equiv\delta S/\delta g_{\mu\nu}$,
$\mathcal{E}^{\mu}\equiv\delta S/\delta A_{\mu}$ evaluated on supersymmetric
configurations. For black-hole-type configurations they are
\cite{Meessen:2006tu}
\begin{equation}
    \begin{array}{rcl}
\mathcal{E}^{0m} & = & \mathcal{E}^{m}
={}^{\star}\mathcal{B}^{m}=
\Im{\rm m}[e^{-i\alpha}(\mathcal{E}^{0}-i{}^{\star}\mathcal{B}^{0})]=0\, ,\\
& & \\
\mathcal{E}^{00} & = &  
\Re{\rm e}[e^{-i\alpha}(\mathcal{E}^{0}-i{}^{\star}\mathcal{B}^{0})]\, .\\
\end{array}
\end{equation}
If we deal with classical solutions, then
$\mathcal{E}^{\mu\nu}=\mathcal{E}^{\mu}=0$ and, where this is not so,
(singularities), we can think on the presence of {\it sources}. Then the KSIs
become constraints for supersymmetric {\it sources}:
\begin{itemize}
\item $\mathcal{E}^{0m}=0\,\,\,{\Rightarrow}$ no {\it sources} of {angular
    momentum} or NUT charge.
\item $\mathcal{E}^{m}={}^{\star}\mathcal{B}^{m}=0 \,\,\,{\Rightarrow}$ no
  {\it sources} of electric or magnetic {dipole momenta}.
\item $\Im{\rm
    m}[e^{-i\alpha}(\mathcal{E}^{0}-i{}^{\star}\mathcal{B}^{0})]\sim
  d^{2}{\omega}= 0 \,\,\,{\Rightarrow}$ no {\it sources} of angular momentum
  or NUT charge.
\end{itemize}
These are precisely the supersymmetric {\it sources} String Theory does not
account for. This excludes automatically the pathological solutions that we
studied in the introduction because all of them correspond to {\it sources} of
angular momentum, NUT charge or dipole momenta.

It should be stressed that Solutions with \textit{global} angular momentum
and dipole momenta are not excluded {\it a priori}, but their sources must be
static electric and magnetic monopoles (Reissner-Nordstr\"om black holes).
This is exactly the opposite to what happens with nuclear magnetic dipole
momenta which always correspond to dipole sources (spin) and not to pairs of
magnetic monopoles \cite{Jackson:1977iu}.  It is, however, possible to show
that in pure $N=2,d=4$ supergravity multipole sources such as those in
Eq.~(\ref{eq:multipole}) satisfying all the supersymmetry constrains will only
give rise to static fields, proving the conjecture made by Hartle and Hawking
\cite{Hartle:1972ya}.

To have solutions with angular momentum we need to add matter fields.


\subsection{$N=2,d=4$ SUGRA coupled to vector multiplets} 

The KSIs are basically identical to those of the pure supergravity theory but
now they contain symplectic-invariant combinations of electric and magnetic
charges and dipole momenta.  Again, one of them is related to the condition
$d^{2}{\omega}=0$ which is the integrability condition of the differential
equation defining $\omega$ \cite{Denef:2000nb,Bates:2003vx}.

Now there is a new KSI involving the equations of motion of the scalars
$\mathcal{E}_{i^{*}}\equiv\delta S/\delta Z^{*\, i^{*}}$:
\begin{equation}
\label{eq:scalarKSI}
\langle\, \mathcal{U}^{*}_{i^{*}}\mid \, \mathcal{E}^{0} \, \rangle  =
{\textstyle\frac{1}{2}} e^{-i\alpha}\mathcal{E}_{i^{*}}\, .    
\end{equation}
This equation means that \textit{when the attractor equations are satisfied},
which is one of the conditions that we require, the l.h.s.~vanishes everywhere
and then, by supersymmetry, the r.h.s.~also vanishes and there are no scalar
sources. It is not known how to account for these sources in String Theory.
  
It is worth giving an explicit example of how the conditions we impose lead in
this context to regular solutions. Let us consider the simple prepotential
$\mathcal{F}=-i\mathcal{X}^{0}\mathcal{X}^{1}$: choose the four real harmonic
functions
\begin{equation}
  \begin{array}{rclrcl}
{\mathcal{I}^{0}} & = & \,\,\,\,\,\,
{\displaystyle
\frac{1}{\sqrt{2}} 
+\frac{q}{r_{1}}+\frac{q}{r_{2}}
}\, ,\hspace{2cm}&
{\mathcal{I}_{0}} & = &  \hspace{1.9cm} 
{\displaystyle-\frac{4{q}}{r_{2}}
}\, ,\\
{\mathcal{I}^{1}} & = &  \,\,\,\,\,\, 
{\displaystyle
\frac{1}{\sqrt{2}} 
+\frac{8{q}}{r_{1}}+\frac{8{q}}{r_{2}}
}\, ,&
{\mathcal{I}_{1}} & = & 
{\displaystyle
-\frac{1}{4\sqrt{2}} 
-\frac{q}{r_{1}}+\frac{q}{r_{2}}
}\, ,\\
\hspace{1cm}r_{1,2} & \equiv & \,\,\,\,\,\, |\vec{x}-\vec{x}_{1,2}|\, ,& & & \\
  \end{array}
\end{equation}
where ${q}>0$. With this choice we get the metric component
\begin{equation}
-g_{rr}= 1 +\frac{9\sqrt{2}q}{r_{1}}  
+\frac{10\sqrt{2}q}{r_{2}}
+\frac{16q^{2}}{r_{1}^{2}}  
+\frac{8q^{2}}{r_{2}^{2}}
+\frac{40q^{2}}{r_{1}r_{2}}\, ,
\end{equation}

\noindent
which is finite everywhere outside $r_{1,2}=0$, the positions of the would-be
\textit{sources} which in the end are regular horizons.  In particular the
``mass'' of each of the two objects is positive
\begin{equation}
M_{1}=9q/\sqrt{2}\, ,
\hspace{.6cm}
M_{2}=5\sqrt{2}q\, ,  
\hspace{.6cm}
M=M_{1}+M_{2}=19q/\sqrt{2}\, .
\end{equation}
In the $r_{1,2}\rightarrow 0$ limits we find spheres of finite areas
\begin{equation}
\frac{A_{1}}{4\pi} = 16q^{2}
= 2|{\mathcal{Z}_{\rm fix,1}}|^{2}\, ,
\hspace{1cm}
\frac{A_{2}}{4\pi} = 8q^{2} 
= 2|{\mathcal{Z}_{\rm fix,2}}|^{2}\, .  
\end{equation}
To have zero NUT charge, we must fix
\begin{equation}
r_{12}\equiv |\vec{x}_{2}-\vec{x}_{1}|=12\sqrt{2}q\, ,   
\end{equation}
and we get a finite \textit{global} angular momentum given by
\begin{equation}
|J|=  12q^{2}\, . 
\end{equation}
The origin of this angular momentum is the angular momentum of the
electromagnetic fields, which is due to the simultaneous presence of electric
and magnetic charges. Therefore, it is naturally quantized.


\subsection{Preliminary results in $N=1,d=5$ SUGRA} 

In this theory the KSIs are qualitatively different, which is to be expected
since in $d=5$ there are regular supersymmetric rotating black holes and black
rings \cite{Breckenridge:1996is,Elvang:2005sa}.  In particular we have a KSI of
the form \cite{kn:BMO}
\begin{equation}
\mathcal{E}^{m0}  =  
-{\textstyle\frac{\sqrt{3}}{4}}h^{I}\mathcal{E}_{I}{}^{m}\, ,
\hspace{.2cm}{\longrightarrow}\hspace{.2cm} 
\text{source of angular and
electric  dipole momenta are related.}
\end{equation}
The metrics of all supersymmetric black holes and rings are of the form
\cite{Gauntlett:2002nw,Gauntlett:2004wh,Gauntlett:2004qy}
\begin{equation}
ds^{2}=f^{2}(dt+{\omega})^{2}
-f^{-1}h_{\underline{m}\underline{n}}dx^{m}dx^{n}\, ,
\end{equation}
and the information about ``NUT charge'' (its $d=5$ equivalent), angular
momentum etc. is, again, contained in the 1-form $\omega$ which is determined
by a differential equation in $d\omega$. The above KSI turns out to be the
integrability of the $\omega$ equation:
\begin{equation}
\mathcal{E}^{m0} +{\textstyle\frac{\sqrt{3}}{4}}h^{I}\mathcal{E}_{I}{}^{m}
={\textstyle\frac{1}{2}}f^{-5/2}
[\star_{4}d^{2}{\omega}]^{m}\, .
\end{equation}
Again, requiring supersymmetry \textit{everywhere} in these 5-dimensional
theories implies that the integrability condition $d^{2}\omega=0$
\textit{everywhere}, but now, as different from what happens in $d=4$, this
does not imply total absence of \textit{sources} of angular momentum, but a
specific relation between these and electric dipole \textit{sources}.  We have
checked that the regular supersymmetric BMPV black hole
\cite{Breckenridge:1996is} and ring \cite{Elvang:2005sa} solutions do satisfy
the integrability condition $d^{2}\omega=0$ \textit{everywhere}.


\section{Final comments}

Some problems have been left aside in the above discussion. 

Why do we impose the attractor equations? Sometimes there is no attractor, but
the KSI Eq.~(\ref{eq:scalarKSI}) is still satisfied. This is typically what
happens for \textit{small black holes}, but in that case they seem to always
satisfy a new quantum-corrected attractor equation and the quantum-corrected
geometry is regular \cite{Dabholkar:2004dq}.

There are other possible pathologies of a supersymmetric solution that do no
seem to be eliminated by our conditions. The most important of them is the
existence of closed timelike curves (CTCs) in some 5-dimensional solutions.
The everywhere regular (it is a Riemannian homogenous space) and maximally
supersymmetric G\"odel solution of $n=1,d=5$ supergravity has CTCs and it is
not clear how the should be interpreted.

Finally, the Calabi-Yau compactifications that give rise to $d=4,5$
supergravities with 8 supercharges give not only vector multiplets but also
hypermultiplets. While it is always consistent to truncate them, these
truncations do not correspond to the generic situation and the modifications
induced by turning on the hyperscalars need to be studied in depth. The first
steps in this direction have been taken in
\cite{Hubscher:2006mr,Bellorin:2006yr}, where the all the supersymmetric
solutions of these ungauged supersymmetric theories with hypermultiplets have
been found. In \cite{Bellorin:2006yr} an asymptotically flat $1/8$
supersymmetric ``deformation'' of the $1/2$ supersymmetric $d=5$
Reissner-Nordstr\"om solution induced by the presence of non-trivial
hyperscalars was constructed and found to be singular\footnote{See the talk by
  P.~Meessen in these proceedings.}.

\section*{Acknowledgements}

The author would like to acknowledge J.~Bellor\'{\i}n and P.~Meessen for their
collaboration and R.~Emparan for useful conversations.  This work has been
supported in part by the Spanish Ministry of Science and Education grant
BFM2003-01090, the Comunidad de Madrid grant HEPHACOS P-ESP-00346 and by the
EU Research Training Network {\em Constituents, Fundamental Forces and
  Symmetries of the Universe} MRTN-CT-2004-005104.




\begin{thebibliography}{99}

\bibitem{Kallosh:1992ii}
R.~Kallosh, A.D.~Linde, T.~Ort\'{\i}n, A.W.~Peet and A.~Van Proeyen,
Phys.\ Rev.\ D {\bf 46}  5278 (1992)
[\hepth{9205027}].

\bibitem{Perjes:1971gv}
Z.~Perj\'es,
Phys.\ Rev.\ Lett.\  {\bf 27} 1668  (1971).

\bibitem{kn:IW} 
W.~Israel and G.A.~Wilson,
J.~Math.~Phys.~\textbf{13}, 865  (1972).

\bibitem{Tod:1983pm}
K.P. Tod,
Phys.\ Lett.\ B {\bf 121} 241  (1983).

\bibitem{kn:Bri} 
D.~Brill,
Phys.\ Rev.\ \textbf{133} B845-B848  (1964).

\bibitem{kn:M}
C.~Misner, J.~Math.~Phys.~\textbf{4} 924  (1963).

\bibitem{Kerr:1963ud}
R.~P.~Kerr,
Phys.\ Rev.\ Lett.\  {\bf 11} 237  (1963).

\bibitem{kn:NCCEPT} 
E. T.~Newman,  E.~Couch, K.~Chinnapared, A.~Exton,
A.~Prakash and R.~Torrence,
J.\ Math.\ Phys.\ \textbf{6} 918 (1965).

\bibitem{Hartle:1972ya}
J.B.~Hartle and S.W.~Hawking,
Commun.\ Math.\ Phys.\  {\bf 26} 87 (1972).

\bibitem{kn:Pa} 
A.~Papapetrou,
Proc.~Roy.~Irish.~Acad.~\textbf{A51} 191 (1947).

\bibitem{Majumdar:1947eu}
S.D.~Majumdar,
Phys.\ Rev.\  {\bf 72} 390 (1947).

\bibitem{kn:BrilLin} 
D.R.~Brill and R.W.~Lindquist,
Phys.~Rev.~\textbf{131} 471-476 (1963).

\bibitem{Behrndt:1997ny}
K. Behrndt, D. L\"ust and W.A. Sabra,
Nucl.\ Phys.\ B {\bf 510} 264 (1998)
[\hepth{9705169}].

\bibitem{Meessen:2006tu}
P.~Meessen and T.~Ort\'{\i}n,
Nucl.\ Phys.\ B {\bf 749} 291 (2006)
[\hepth{0603099}].

\bibitem{Bellorin:2006xr}
J.~Bellor\'{\i}n, P.~Meessen and T.~Ort\'{\i}n,
\hepth{0606201}.

\bibitem{Kallosh:1993wx}
R.~Kallosh and T.~Ort\'{\i}n,
\hepth{9306085}.

\bibitem{Bellorin:2005hy}
J.~Bellor\'{\i}n and T.~Ort\'{\i}n,
Phys.\ Lett.\ B {\bf 616} 118 (2005)
[\hepth{0501246}].

\bibitem{Ferrara:1995ih}
S.~Ferrara, R.~Kallosh and A.~Strominger,
Phys.\ Rev.\ D {\bf 52} 5412 (1995)
[\hepth{9508072}].

\bibitem{Strominger:1996kf}
A.~Strominger,
Phys.\ Lett.\ B {\bf 383} 39 (1996)
[\hepth{9602111}].

\bibitem{Ferrara:1996dd}
S. Ferrara and R. Kallosh,
Phys.\ Rev.\ D {\bf 54} 1514 (1996)
[\hepth{9602136}].

\bibitem{Ferrara:1996um}
S.~Ferrara and R.~Kallosh,
Phys.\ Rev.\ D {\bf 54} 1525 (1996)
[\hepth{9603090}].

\bibitem{Jackson:1977iu}
J.D.~Jackson,
Yellow Report CERN-77-17
\href{http://www.slac.stanford.edu/spires/find/hep/www?r=cern-77-17}{SPIRES entry} 
Reprinted in V.~Stefan and V.F.~Weisskopf, eds., Physics and Society:
Essays in Honor of Victor Frederick Weisskopf (AIP Press, New York, Springer,
Berlin, 1998) 236pp.

\bibitem{Denef:2000nb}
F.~Denef,
JHEP {\bf 0008} 050 (2000)
[arXiv:\hepth{0005049}].

\bibitem{Bates:2003vx}
B.~Bates and F.~Denef,
\hepth{0304094}.

\bibitem{Breckenridge:1996is}
J.C.~Breckenridge, R.~C.~Myers, A.~W.~Peet and C.~Vafa,
Phys.\ Lett.\ B {\bf 391} 93 (1997)
[\hepth{9602065}].

\bibitem{Elvang:2005sa}
H.~Elvang, R.~Emparan, D.~Mateos and H.S.~Reall,
JHEP {\bf 0508} 042 (2005)
[\hepth{0504125}].

\bibitem{kn:BMO} 
J.~Bellor\'{\i}n, P.~Meessen and T.~Ort\'{\i}n, (to appear).

\bibitem{Gauntlett:2002nw}
J.~P.~Gauntlett, J.~B.~Gutowski, C.~M.~Hull, S.~Pakis and H.~S.~Reall,
Class.\ Quant.\ Grav.\  {\bf 20} 4587 (2003)
[\hepth{0209114}].


\bibitem{Gauntlett:2004wh}
J.P.~Gauntlett and J.B.~Gutowski,
Phys.\ Rev.\ D {\bf 71} 025013 (2005)
[\hepth{0408010}].

\bibitem{Gauntlett:2004qy}
J.~P.~Gauntlett and J.~B.~Gutowski,
Phys.\ Rev.\ D {\bf 71} 045002 (2005)
[\hepth{0408122}].

\bibitem{Dabholkar:2004dq}
A.~Dabholkar, R.~Kallosh and A.~Maloney,
JHEP {\bf 0412} 059 (2004)
[\hepth{0410076}].

\bibitem{Hubscher:2006mr}
M.~H\"ubscher, P.~Meessen and T.~Ort\'{\i}n,
\hepth{0606281}.

\bibitem{Bellorin:2006yr}
J.~Bellor\'{\i}n, P.~Meessen and T.~Ort\'{\i}n,
\hepth{0610196}.


\end{thebibliography}
\end{document}